\documentclass[11pt,a4paper]{article}
\usepackage{geometry,graphicx}
\title{Extended AIGER Format for Synthesis}
\author{Swen Jacobs\\[1em] {\sf swen.jacobs@iaik.tugraz.at}}

\begin{document}
\pagestyle{empty}
\setcounter{secnumdepth}{2}
\maketitle

\begin{abstract}
We extend the AIGER format, as used in HWMCC, to a format that is
suitable to define synthesis problems with safety specifications.
We recap the original format
and define one format for posing synthesis problems and one for
solutions of synthesis problems in this setting. 
\end{abstract}

\section{Original Format}

The AIGER format was introduced by Biere~\cite{aiger}, and is used as
the standard format in the hardware model checking competition
(HWMCC)~\cite{HWMCC}. We use version \texttt{20071012} of the format. 
A file in AIGER format (ASCII variant) consists of the following parts:
\begin{enumerate}
\item Header
\item Input definitions
\item Latch definitions
\item Output definitions
\item And-gate definitions
\item Symbol table (optional)
\item Comments (optional)
\end{enumerate}

\subsubsection{Header}
The header consists of a single line

\texttt{ aag M I L O A }

\noindent where 
\texttt{M} gives the maximum variable index, and \texttt{I, L, O, A} the number of inputs, latches, outputs and AND-gates, respectively. In the following, all numbers that represent inputs, latches, outputs or AND-gates need to be not greater than $2$\texttt{M}$+1$.

\subsubsection{Input definitions}

Every input definition takes one line, and consists of a single even number. (Even numbers represent Boolean variables, an odd number $n+1$ represents the negation of $n$. Inputs are never directly negated, so they are always represented by even numbers.)

\subsubsection{Latch definitions}

Every latch definition takes one line, and consists of an even number (defining the variable that represents the latch), followed by a number that defines which variable is used to update the latch in every step. By default, latches are assumed to have initial value $0$. With an extension (see Section~\ref{sec:extensions}), it is possible to explicitly define the initial value of a given latch.

\subsubsection{Output definitions}

Every output definition takes one line, and consists of a single
number (representing a possibly negated input, latch, or
AND-gate). For our class of (safety) problems, there is exactly one
output, and safety conditions are encoded such that the circuit is
safe if the output is always $0$.

\subsubsection{AND-gate definitions}

Every AND-gate definition takes one line, and consists of three numbers. The first is an even number, representing the output of the AND-gate, and is followed by two numbers representing its (possibly negated) inputs.

\subsubsection{Symbol table}

The symbol table assigns names to inputs, latches and outputs. It is
optional, and need not be complete. Every line defines the name of one
input, latch, or output, and starts with \texttt{i,l,o}, respectively,
followed by the number of the input, latch, or output in the sequence
of definitions (\emph{not} the variable index of the input - so the
first input is always defined by a line starting with \texttt{i0}, the
first latch with \texttt{l0}). This is followed by an arbitrary string
that names the variable. 

\section{Modified AIGER format for synthesis specifications}
\label{sec:synthesis-input}

We propose a simple way to extend the AIGER format for controller
synthesis: reserve the special string ``\texttt{controllable\_}'' in
the symbol table, and prepend it to the names of controllable input
variables. All other input variables are implicitly controlled by the
environment.

In this way, we can turn benchmarks from HWMCC into synthesis
benchmarks by defining a subset of the inputs to be controllable. We
may even keep an existing symbol table and only modify
the names of controllable variables, prepending this string. 

\section{Output of synthesis tools in AIGER format}

Starting from an input as defined in
Section~\ref{sec:synthesis-input}, we define when an AIGER file is a
solution of the synthesis problem. We do this in such a way that
the solution contains the specification circuit, and can directly be
checked by existing model checkers that support the AIGER format.

\subsection{Syntactical correctness}

Up to the exceptions noted below, the output file must contain all lines of the input file, in the same order.

\subsubsection{Header} The header line must be modified to 

\texttt{ aag M' I' L' O A' }

where 

\begin{itemize}
\item \texttt{I'} $=$ \texttt{I} $-\ c$ (for $c$ controllable inputs in the specification)
\item \texttt{L'} $=$ \texttt{L} $+\ l$ (for $l$ additional latches defined in the controller)
\item \texttt{A'} $=$ \texttt{A} $+\ a$ (for $a$ additional AND-gates defined in the controller)
\item \texttt{M'} $=$ \texttt{I'} $+$ \texttt{L'} $+$ \texttt{A'}
\end{itemize}

The correct value for $c$ can be computed from the symbol table of the input file, while correct values for $l$ and $a$ need to be computed from the output file (the number of lines that define new latches or AND-gates, respectively).

\subsubsection{Inputs}
Definitions for uncontrollable inputs remain unchanged. Definitions for controllable inputs are removed, and the corresponding variable indices have to be redefined either as new latches or AND-gates (see below).

\subsubsection{Latches}
No definitions of latches may be removed, but additional latches may be defined in the lines following the original latches.

\subsubsection{Outputs}
No definitions of outputs may be removed, no additional outputs may be defined.

\subsubsection{AND-gates}
No definitions of AND-gates may be removed, but additional AND-gates may be defined in the lines following the original AND-gates.

\subsubsection{Global restrictions}
All indices of controllable inputs have to be redefined exactly once, as either a new latch or a new AND-gate. New latches and AND-gates may be defined using the remaining (non-controllable) inputs, any latches, or newly defined AND-gates, but \emph{not} original AND-gates.\footnote{The reason for disallowing original AND-gates is that we want the controller to work only based on the \emph{state} of the given circuit (represented by values of latches), and the uncontrollable inputs. This makes it easier to check correctness of the controller, as we cannot have combinational loops. Note that original AND-gates can and will be duplicated in the controller, if necessary.}

\subsubsection{Symbol table}
The symbol table remains unchanged.

\subsection{Semantical Correctness}

All input files will have the same structure as safety specifications used in HWMCC. In particular, this means that there is only one output, and the system is safe if and only if this output remains $0$ for any possible input sequence.

Any output file satisfying the syntactical restrictions above is an AIGER file itself. It is \emph{correct} if for any input sequence (of the uncontrollable inputs), the output always remains $0$. We say that it is a \emph{solution} to the synthesis problem if it is successfully model checked by an AIGER-capable model checker within a determined time bound.

\section{Other Options}

We have discussed several other options for the format, but currently
decided against them. These are:

\subsubsection{Latches cannot directly be used}
If we forbid reuse of latches in the synthesized component, the
synthesis tool must decide which parts of the original circuit must be
rebuilt, and pay the price for it. I.e., this variant allows for an
additional metric that distinguishes synthesis tools wrt. to the size
of the overall circuit (instead of only the controller). 

\subsubsection{AND-gates can directly be used}
Another variant is to allow original AND-gates to be used in
definitions of new latches or AND-gates. It allows smaller solutions,
but the danger here is that a combinational loop is constructed.

\subsubsection{Output file only contains new or modified parts}
Instead of integrating the new and modified parts into the input file,
we may construct an output that only contains these parts, and is
automatically combined with the input file to get a checkable
AIGER-file. This shifts some of the burden from the synthesis tool to
the checker, but does not seem to be a significant difference. 

\section{Possible Extensions}
\label{sec:extensions}

As a preparation to the upcoming AIGER format version 2.0, the current
format has been extended with some new features~\cite{aiger}, while
staying downward compatible to the basic format described above. In
the future, we could also allow these extensions in the synthesis
format.

\subsubsection{Reset Logic}

A latch can be initialized by adding another literal to the line that defines it. This literal can be $0$ (the default), $1$, or the literal of the latch itself, which means it starts uninitialized.

\subsubsection{Multiple Properties}

To allow using and checking multiple properties for the same model, the header can be extended with additional decimal numbers \texttt{B C J F}, giving the number of \emph{bad state} properties, invariant constraints, justice properties, and fairness constraintes, respectively. The body of the file is then extended with according descriptions of these properties, as defined in the original description of AIGER 1.9~\cite{aiger}.

\bibliographystyle{plain}
\bibliography{local}

\begin{thebibliography}{1}

\bibitem{aiger}
Armin Biere.
\newblock {\em {AIGER} Format and Toolbox}.
\newblock {\tt http://fmv.jku.at/aiger/}.

\bibitem{HWMCC}
Armin Biere.
\newblock {\em Hardware Model Checking Competition}.
\newblock {\tt http://fmv.jku.at/hwmcc/}.

\end{thebibliography}

\end{document}